\documentclass[12pt]{iopart}
\usepackage{graphicx}

\begin{document}

\title[Low coherency of wind induced seismic noise]{Low coherency of wind induced seismic noise: Implications for gravitational wave detection}

\author{ H. Satari$^1$, C. Blair$^1$, L. Ju$^1$, E. Saygin$^{1,2}$, D. Blair$^1$, C. Zhao$^1$, D. Lumley$^3$, Patrick Meyers$^4$}

\address{$^1$ School of Physics, University of Western Australia, Australia}
\address{$^2$ CSIRO, Deep Earth Imaging Future Science Platform, Australia}
\address{$^3$ Dept Geosciences, Dept Physics, University of Texas at Dallas, USA}
\address{$^4$ School of Physics, University of Melbourne, Australia}
\ead{hamid.satari@research.uwa.edu.au}
\vspace{10pt}
\begin{indented}
\item[]May 2022
\end{indented}

\begin{abstract}
Seismic noise poses challenges for gravitational wave detection. Effective vibration isolation and methods to subtract unsheildable Newtonian Noise are examples. Seismic arrays offer one way to deal with these issues assuming seismic coherency. In this paper we find that wind induced seismic noise is incoherent and will dramatically reduce the projected low frequency sensitivity of future gravitational wave detectors. To quantify this, we measure the coherence length of wind induced seismic noise from 0.06--20~Hz in three distinct locations: close to a building, among tall trees and in shrubs. We show that wind induced seismic noise is ubiquitous and reduces the coherence lengths form several hundred meters to 2--40~m for 0.06--0.1~Hz, from $>$60~m to 3--16~m for 1.5--2.5~Hz and from $>$35~m to 1--16~m around 16.6 Hz frequency bands in the study area. This leads to significant loss of velocity and angular resolution of the array for primary microseism, 5 times worse Newtonian Noise cancellation by wiener filtering at 2~Hz, while it does not pose additional challenge for Newtonian Noise cancellation between 10--20~Hz.

\end{abstract}
\section{Introduction}

Isolation of next generation gravitational wave (GW) detectors from ambient seismic noise is one of the fundamental instrumental challenges for future detectors especially at frequencies below 20~Hz. Seismic arrays have been widely proposed  \cite{coughlin2014,Driggers, Coughlin_2016,Badaracco2019} to actively assist vibration isolation systems by a feed-forward control \cite{Feedforward,Derossa2012} and cancel the ground motion before it enters the system \cite{Abbott_2004}. Arrays can also be exploited to characterize the ambient seismic noise sources which improves prediction and subtraction of Newtonian Noise (NN) \cite{coughlin2014, Badaracco2019, Badaracco2020, Driggers, Coughlin2016}. These techniques, and generally any array seismology method rely on the coherence of signals between array elements \cite{Coughlin_2019}. Therefore, incoherent seismic noise that can mask coherent ambient seismic noise reduces the capability of the seismic array to measure the properties of the coherent portion of the seismic wavefield. 

Sensitivity of seismic measurements to wind has been extensively studied. References \cite{Johnson2019, Zieger_2018} demonstrated the coherent signature of wind induced seismic noise, while references \cite{withers_1996, Dybing_2019} reported low coherency of this type of seismic noise though with limited and sparse spatial measurements. By comparing wind speed and direction from two anemometers, \cite{Dybing_2019} showed that wind is incoherent on the 100~m length scale. Analyzing the spectral contribution to the seismic field from wind turbines in the vicinity of VIRGO \cite{virgo_turbine} showed an example of coherent wind induced seismic noise that impacts a gravitational wave detector. The turbulent interaction of wind with surface objects that results in the more challenging low coherent wind induced seismic noise is the subject of this paper. We use higher sensor density than \cite{withers_1996, Dybing_2019} and measure coherence lengths to analyze the coherency of wind induced seismic noise. We use the ability to resolve background coherent seismic noise to highlight the effect. This allows us to quantify the effect of wind induced seismic noise on GW detection assuming Wiener filters are used for subtraction of an error signal derived from the array. 

The analysis focuses on 0.06--20~Hz which encompasses main seismic noises that matters for advanced GW detection (microseism and NN). In section 2, we present the results of a pilot study using a simple seismic array at Gingin High Optical Power Facility (HOPF) in Western Australia (WA). The spectral analyses show strong correlation between ground motion and wind speed for a significant portion of the seismic spectrum. Coherence length measurements indicate three frequency bands below 20~Hz that suffer from low coherency of wind induced seismic noise. This information guided the configuration of the dense seismic array for the purpose of measuring the coherence length of wind induced seismic noise. In section 3, we present the results of the coherence length measurements from the dense seismic array. Finally in Section 4, we discuss the implications of the low coherency of wind induced seismic noise on next generation GW detectors. 

\begin{figure}
\centering
\includegraphics{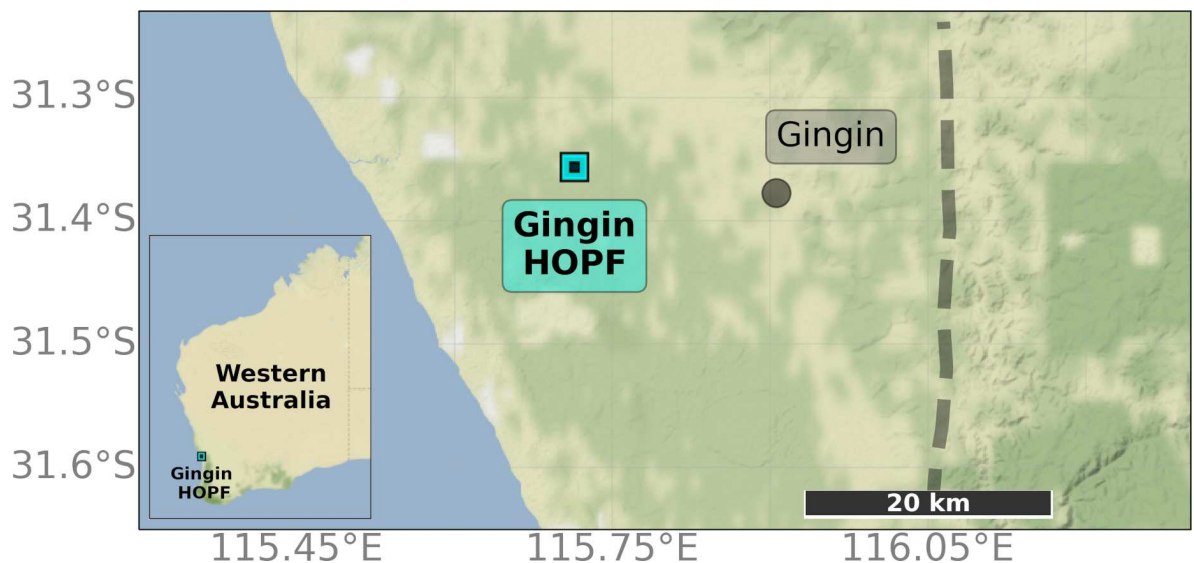}
\caption{Location of the study area in Gingin, Western Australia on Google Earth image with the inset map showing that on Western Australia. The dashed line shows the Darling fault.}
\label{Figure_1}
\end{figure}

\section{Information from the pilot study}

\subsection{Study area}

HOPF is a research center working on harnessing high optical power to increase the sensitivity of laser interferometer GW detectors \cite{zhao}. It is located in Yeal Natural Reserve near Gingin, 71 km north of the Perth city centre, 17 km to the east of the WA coast and 33 km to the west of the Darling fault which is a relatively inactive faultline \cite{geosciences9100408} that extends for almost 1000 km parallel to the WA coast (figure~\ref{Figure_1}). For its proximity to the WA coast, the study area is subject to large amplitude coherent microseism between 0.06--1~Hz \cite{Nishida2017}. Microseism is coherent over $\approx 5km$ at 0.2~Hz \cite{Coughlin_2019}. This justifies apertures of Km scale for the spatial sampling of their large wavelengths to improve angular and velocity resolution \cite{Wathelet2008}. However, we do not need such a large aperture size to study the impact of wind on seismic coherency owing to the absence of regional roughness and topographically induced wind gusts known as topography multipliers \cite{Schofield}. HOPF is within a vast zone of remnant coastal belt sand dunes and associated shoreline deposits \cite{vehicle} with maximum 50 meters variation in altitude over 10 km. This flatness allows the study of the interaction of wind with local surface objects like trees and buildings without confounding effects from large valleys or hills. The uniformity of the area topology allows generalization of the results over the entire area. The study area is located in semi-arid bush land composed of shrubs and lightly forested areas. There are three research lab buildings in the study area (the Central-Station, South End-Station and the East End-Station), housing two 80 m long suspended cavities (inset map in figure~\ref{Figure_3}(a)).

\subsection{Preliminary wind induced seismic noise characterization}

\begin{figure}
    \centering
    \includegraphics{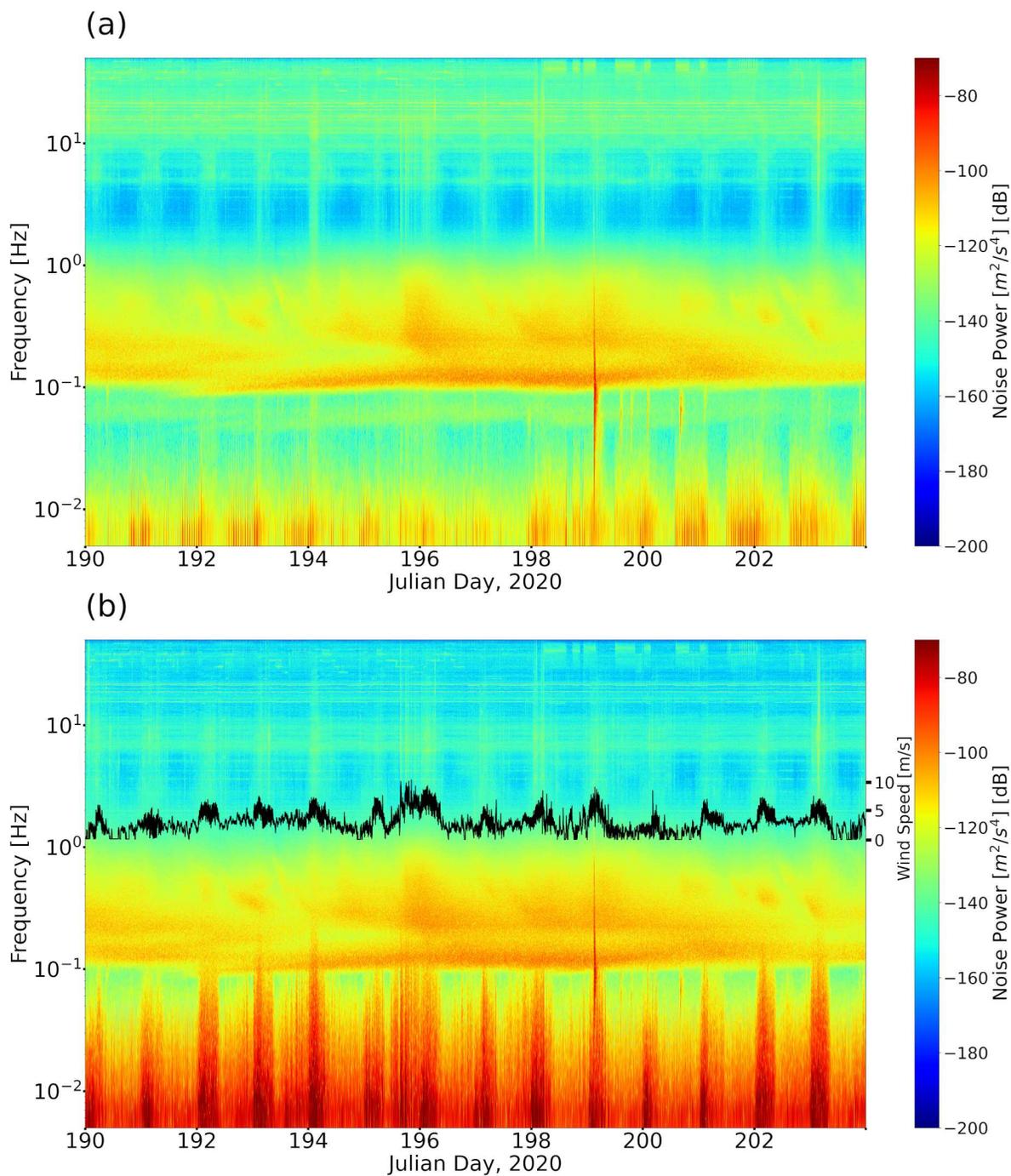}
    \caption{Spectrogram for (a) vertical and (b) horizontal (North-south) components of ground motion recorded at a seismometer in the Central-Station building for days 190--204 of 2020. Scaled wind speeds (black) recorded at the roof of the same building are overlaid on the horizontal component. Wind induced seismic noise is evident in the form of a diurnal variation of the power correlated wind speed across much of the spectrum below 0.1~Hz and above 1~Hz especially on the horizontal component.}
    \label{Figure_2}
\end{figure}

Motivated by the need of a seismic characterization of a GW observatory site (\cite{LIGO_site}, \cite{VIRGO_site}) We conducted a pilot study of the coherent ambient seismic noise for frequencies below 10 Hz using beamforming\cite{Kelly1967, Rost2002}. The seismometers used were three-component Trillium Compact (Nanomterics) and the data was recorded at 4 millisecond sampling rate. An interesting observation from this study was that the beam-power, defined as the absolute coherent component, was considerably
lower each day in a period where single seismometer amplitudes were high. The diurnal seismic noise was strongly correlated with wind speed and found across a wide spectrum. The spectrogams in figure~\ref{Figure_2} show this for two weeks of the data acquired in the Central-Station building during July 2020. Scaled wind speeds are overlaid in black for a visual correlation.

The vertical component spectrogram (figure~\ref{Figure_2}(a)) better shows the underlying dominant seismic noise e.g. the primary, secondary, and tertiary microseism peaks around 0.07~Hz, 0.15~Hz and 0.5~Hz respectively. On the other hand, the horizontal component spectrogram (figure~\ref{Figure_2}(b)) more clearly illustrates the daily power increase in the first half of the Julian days, equating to 8am to 8pm local time. Strong wind induced seismic noise in the horizontal component can be seen from the correlation between seismic power and wind speed at frequencies below 0.1~Hz and above 1~Hz. It has contaminated other seismic background noise in several frequency bands, particularly the primary microseisem peak at 0.07~Hz which is mostly visible in the vertical component but hidden by the diurnally varying noise in the horizontal component. The sharp onset of a magnitude 7.0 earthquake in the initial hours of day 199 is another example of how local wind induced seismic noise can mask coherent seismic signals. This event can be seen on both vertical and horizontal channels. However, in the vertical component it is the dominant signal between 0.01~Hz and 3~Hz, while in the horizontal component its low frequency portion (0.01--0.07~Hz) is masked by wind induced seismic noise.

\begin{figure}
    \centering
    \includegraphics{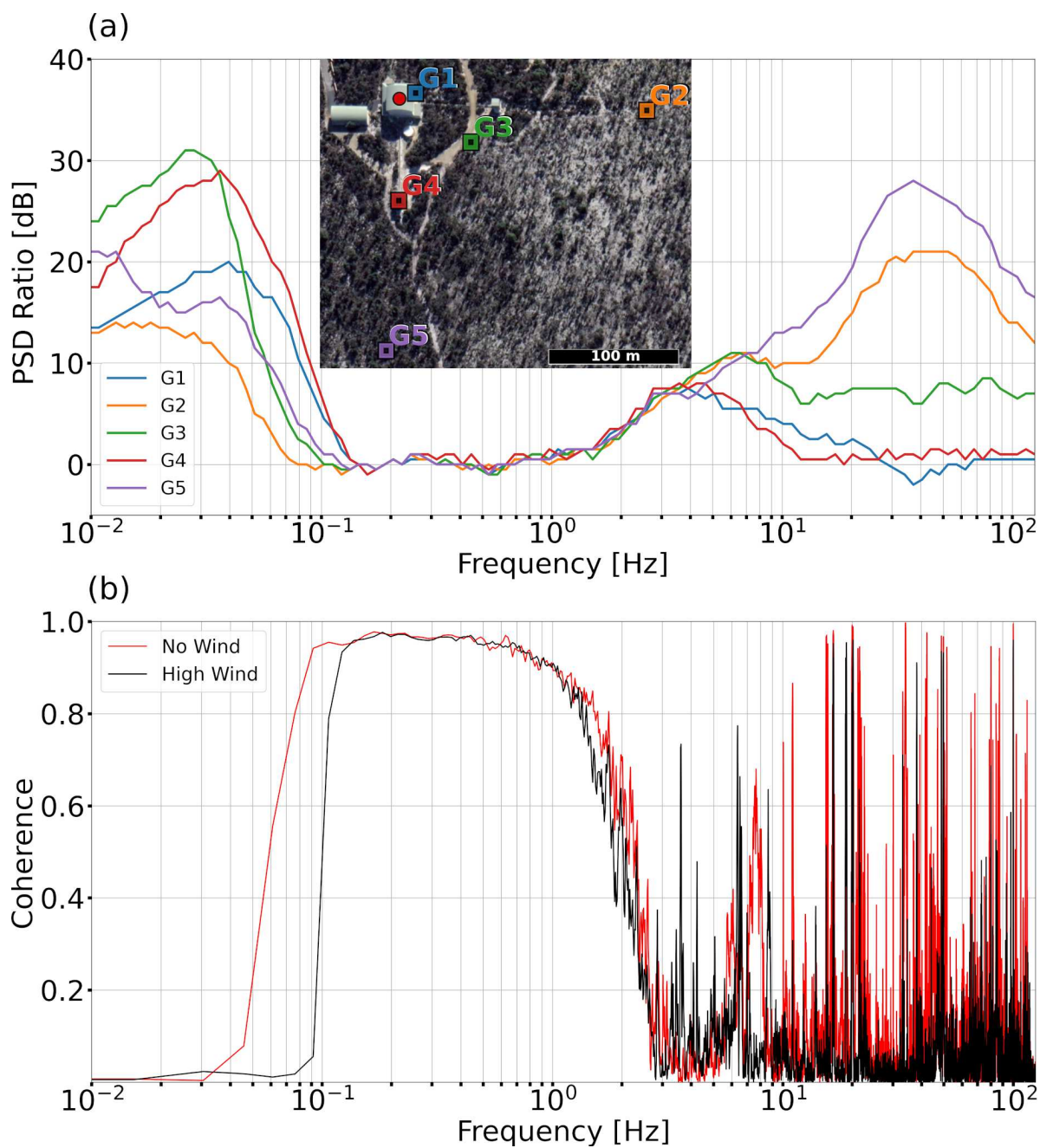}
    \caption{Effect of wind on seismic data across the whole seismic spectrum throughout the study area. (a) PSD ratios calculated by dividing horizontal seismic PSD during 2 windy hours by that during 2 non-windy hours. The inset map shows the location of each seismometer with the corresponding color in the pilot study. (b) Coherence of horizontal seismic data between two seismometers 60 m apart from one another (G1 and G3 in the inset map shown in (a)). These measurements were made during the same non-windy (red line) and windy hours (black line) as those chosen for the PSD ratios in (a).}
    \label{Figure_3}
\end{figure}

To demonstrate the effect of wind induced seismic noise throughout the site, the ratios of windy to non-windy seismic power spectral density (PSD) for all of the seismometers used in the pilot study is shown in figure~\ref{Figure_3}(a). G1 and G4 are inside the Central-Station (main laser building) and South End-Station (suspended end mirror location of the 80 m cavity arm) respectively. The other three seismometers were buried 1 m deep in the ground outside the buildings. The PSD ratios show that wind induced seismic noise is dominant over other seismic noise sources across almost the entire seismic spectrum except the peak of microseism (between 0.1--1~Hz) which is consistent with the spectrograms of the G1 seismometer shown in figure \ref{Figure_2}. The coherence for the horizontal component between the two closest seismometers, G1 and G3 are provided and compared during the same windy and non-windy hours in figure~\ref{Figure_3}(b). 

PSD ratios are considerably high below 0.1~Hz for all of the locations which has caused a significant decrease in the coherence from 0.06~Hz to 0.1~Hz (primary microseism) over only 60 m (figure~\ref{Figure_3}(b)). This demonstrates the effect of wind induced seismic noise on GW detectors: where there is a coherent seismic noise that can be measured using a seismic array to be addressed via a feed-forward control, wind reduces coherency by injecting a relatively incoherent noise that prevents the array from efficiently measuring the background coherent noise. Another frequency band is 1.5--2.5~Hz with increased PSD ratios and decreased coherence. We had previously observed the resulted degrading effect of wind on the array's performance from beamforming analysis for this frequency band.

It may be noted that unlike the above mentioned frequency bands, there is an increase in the coherence between 3--7~Hz with increased wind. This is related to a 40~m tall leaning tower in the north-west of the Central-Station 300~m apart from it. This structure has several resonant modes in this frequency band that are driven by wind and generates coherent seismic waves. While interesting in its own right, in this paper we will focus on the incoherent wind driven noise. For frequencies above 4~Hz, we see a reduced effect of wind induced seismic noise for G1 and G4 in their PSD ratios. These two seismometers are mounted on the concrete slabs which attenuated the high frequency wind effect inside the laboratory buildings.

There is little evidence for the hypothesis that interactions of surface structures with wind create seismic disturbances in this data. From heights to lowest PSD ratios below 0.1 Hz (figure~\ref{Figure_3}(b)), the five seismometers were surrounded by fences (G3), a small building (G4), a larger and taller building (G1), dense bush and tall trees (G5), sparse shrubs and short trees (G2). Although there is a large location dependency in the PSD ratios below 0.1 Hz, it does not appear to follow a rule based on surface structures. Above 1~HZ there is some evidence. For example trees make more seismic disturbance than shrubs. In
order to better quantify the effect of low coherency of wind induced seismic noise on seismic measurements and its location dependence we designed a dedicated array.

\section{Coherence length measurements}

\begin{figure}
    \centering
    \includegraphics{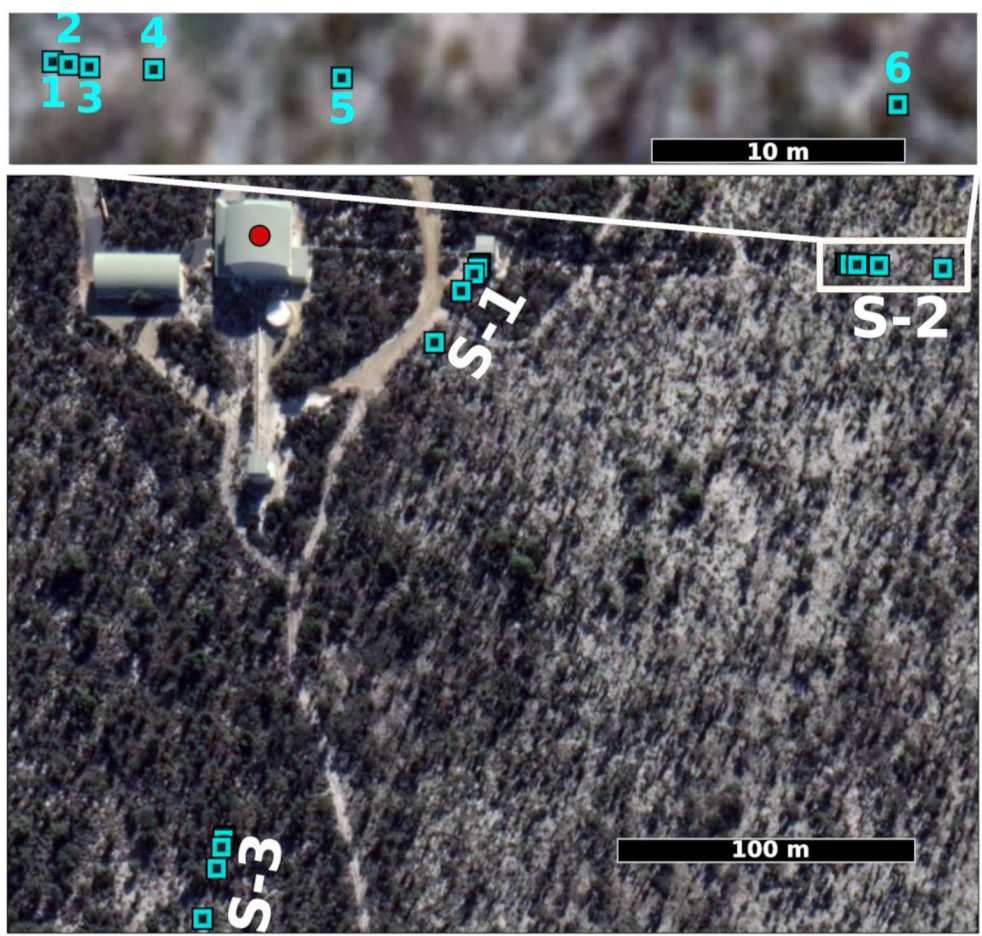}
    \caption{Locations of 18 seismometers in the dedicated array designed to study the coherency of wind induced seismic noise in Gingin HOPF on Google Earth image. There are three groups of six nodes arranged in linear sub-arrays with logarithmic separation in three different locations near a building (S-1) shrubs (S-2) and trees (S-3). The inset map shows the logarithmic separations between the nodes in sub-array S-2. The other 2 sub-arrays are similar. The red circle on top of the Central-Station building shows the location of the anemometer.}
    \label{Figure_4}
\end{figure}

To investigate the interaction of different surface objects with wind, we deployed three sub-arrays in distinct locations, each composed of six broadband seismometers (figure~\ref{Figure_4}). S-1 was chosen as a location close to the East End-Station with seismometers buried outside the building. It is seismically quieter at frequency band above 5~Hz compared to the Central-Station building. The other two sub-arrays, S-2 and S-3, are in shrubs and amongst tall trees respectively. Having already observed the low coherence of wind induced seismic noise for low frequency (0.06--1~Hz) over only 60 m in the pilot study (figure~\ref{Figure_3}(b)), we chose a shorter maximum sensor separation for the sub-arrays with logarithmic spacing for a wide range of coherence length measurements. Therefore, each sub-array has 0.5~m as the minimum, and 35~m as the maximum separation between seismometers. The aim was to capture the transition of the coherent wind induced seismic noise to incoherent wind induced seismic noise.

We use the sensor-sensor coherence between seismometers $i, j$

\begin{equation}
    \gamma_{i,j} (f) = \frac{ \langle x_i(f)x_j^*(f) \rangle}{\sqrt{\langle |x_i(f)|^2\rangle \langle |x_j^*(f)|^2\rangle}},
    \label{eq1}
\end{equation} where $x_i(f)$ is the complex value of the Fourier Transform at a particular frequency $f$ for the \textit{i}th seismomter, $x_i^*(f)$ is its complex conjugate, and $\langle\rangle$ indicate an average over consecutive time windows. By calculating coherence for different pairs of sensors in the dense array and plotting coherence against sensor separations we obtain coherence length by finding on the plot the sensor separation where the coherence drops below 0.5.

\begin{figure}
    \centering
    \includegraphics{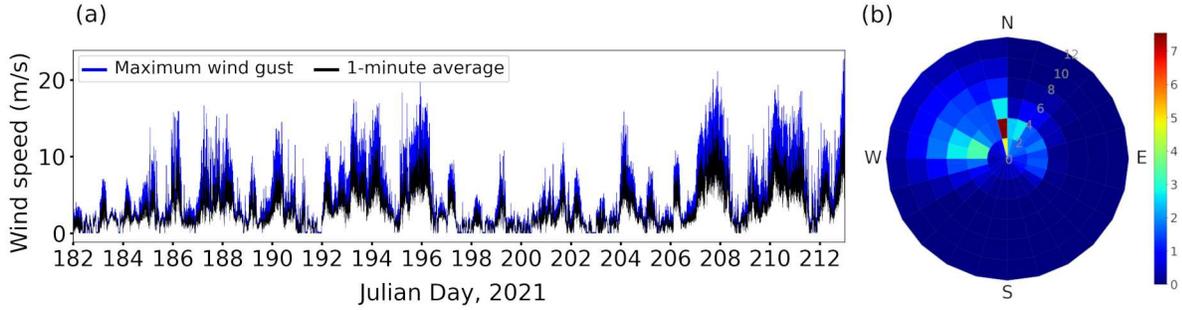}
    \caption{(a) 1 minute averaged wind speed (black) and maximum wind gusts for the same 1 minute windows (blue) during July 2021. (b) 2D histogram of the speed and direction of the wind data where the color bar shows the histogram count. The radial coordinate is wind speeds with 2 m/s intervals marked on the plot. The angular coordinate is the wind direction while N points to the north of the study area. Wind is dominantly coming from west and north marked by W and N respectively.}
    \label{Figure_5}
\end{figure}

For the dedicated wind induced seismic noise study, we used an anemometer (Intech WS3-WD-TB-CL) to measure wind speeds during 2 weeks in July 2021 and processed them into 1 minute averages and maximum values (figure~\ref{Figure_5}(a)). The average wind speed for this period is 3.86 m/s with maximum wind gust reaching 23.49 m/s. Figure~\ref{Figure_5}(b) shows the 2D histogram of the wind speed and wind direction which is predominantly between the West and the North. Wind speeds between 2--4 m/s coming from the North account for the highest joint percentage ($<$ 7\%), while the majority of wind with speeds greater than 4 m/s come from the West and North-west. 

To assess the impact of wind we will compare the coherence length in a windy and non-windy period. The windy period is hours 4--6 of day 210 with 9.7~m/s average wind speed in an azimuth between 320 and 330 degrees which is representative of the general direction of the high wind speeds in the study area during the study period. The non-windy period chosen was hours 22-24 of day 202, with average wind speed of 0.1 m/s.

\begin{figure}
    \centering
    \includegraphics{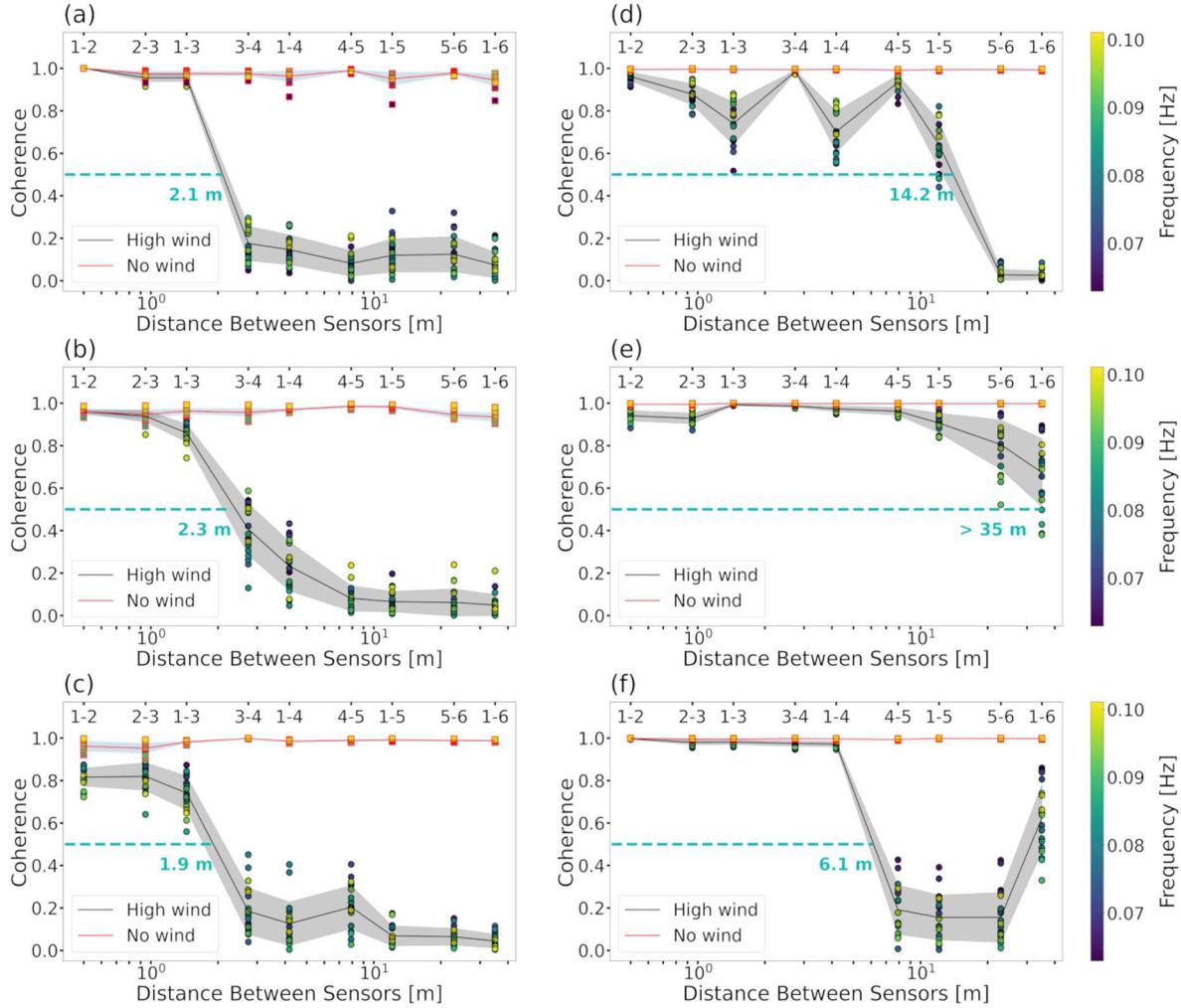}
    \caption{Bandlimited (0.06--0.1~Hz) coherence vs distance between pairs of seismometers for the three sub-arrays in figure~\ref{Figure_4}. High wind (black) is compared to non-windy (red) for horizontal components in  S-1 (a) (building), S-2 (b) (shrubs) and S-3 (c) (trees), then similarly for vertical components in (d), (e) and (f) respectively. Sensor combinations are indicated on top of the plots. Each dot is the coherence measured for the frequency indicated by color. Solid lines are the averages over the frequency band and shaded areas represent one standard deviation. The shortest lengths with a coherence of 0.5 are marked using the blue dashed lines.}
    \label{Figure_6}
\end{figure}

\begin{figure}
    \centering
    \includegraphics{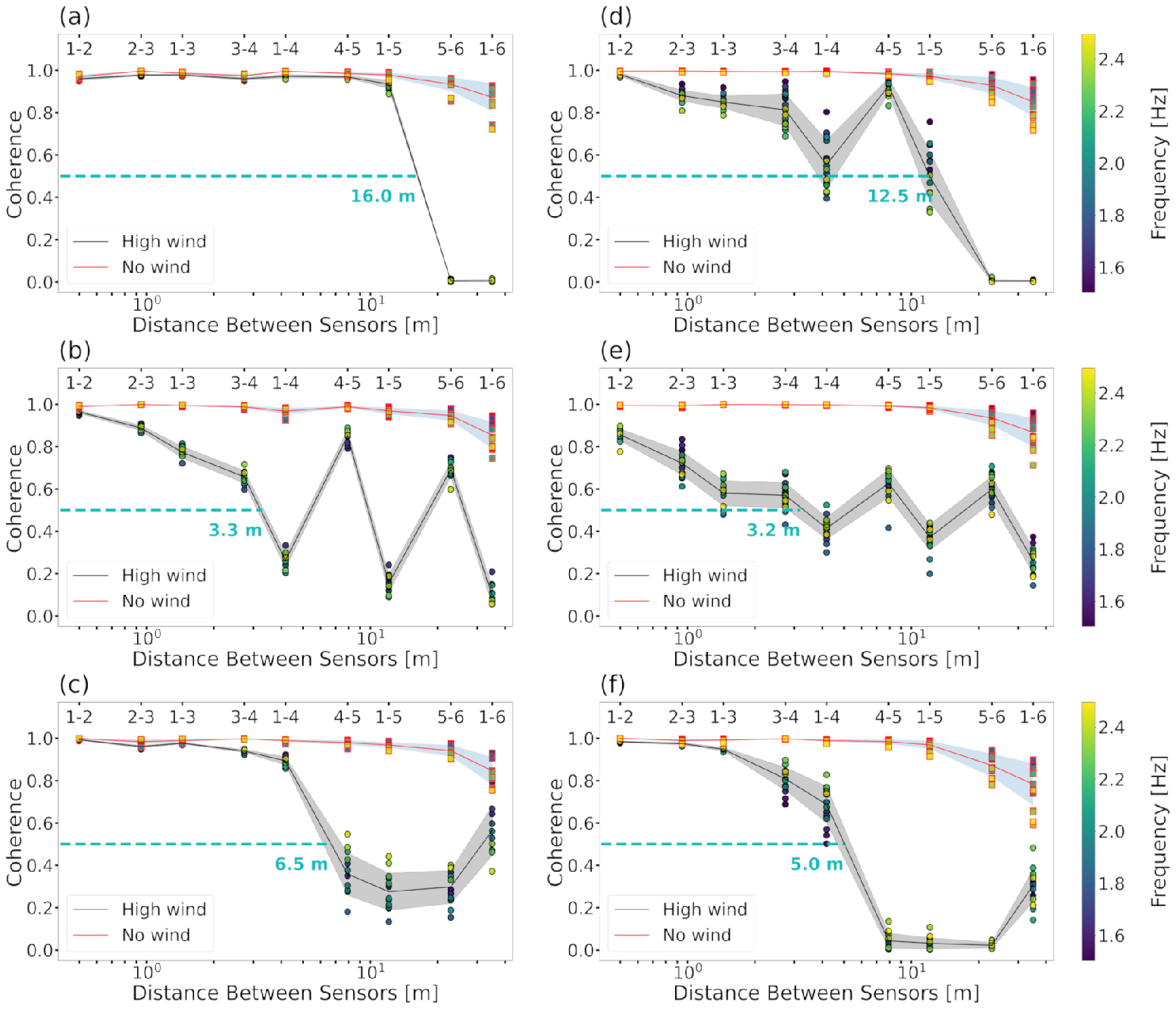}
    \caption{Similar to figure~\ref{Figure_6} but for 1.5--2.5~Hz frequency band. High wind (black) is compared to non-windy (red) for horizontal components in  S-1 (a) (building), S-2 (b) (shrubs) and S-3 (c) (trees), then similarly for vertical components in (d), (e) and (f) respectively. Solid lines are the averages over the frequency band and shaded areas represent one standard deviation.}
    \label{Figure_7}
\end{figure}

\begin{figure}
    \centering
    \includegraphics{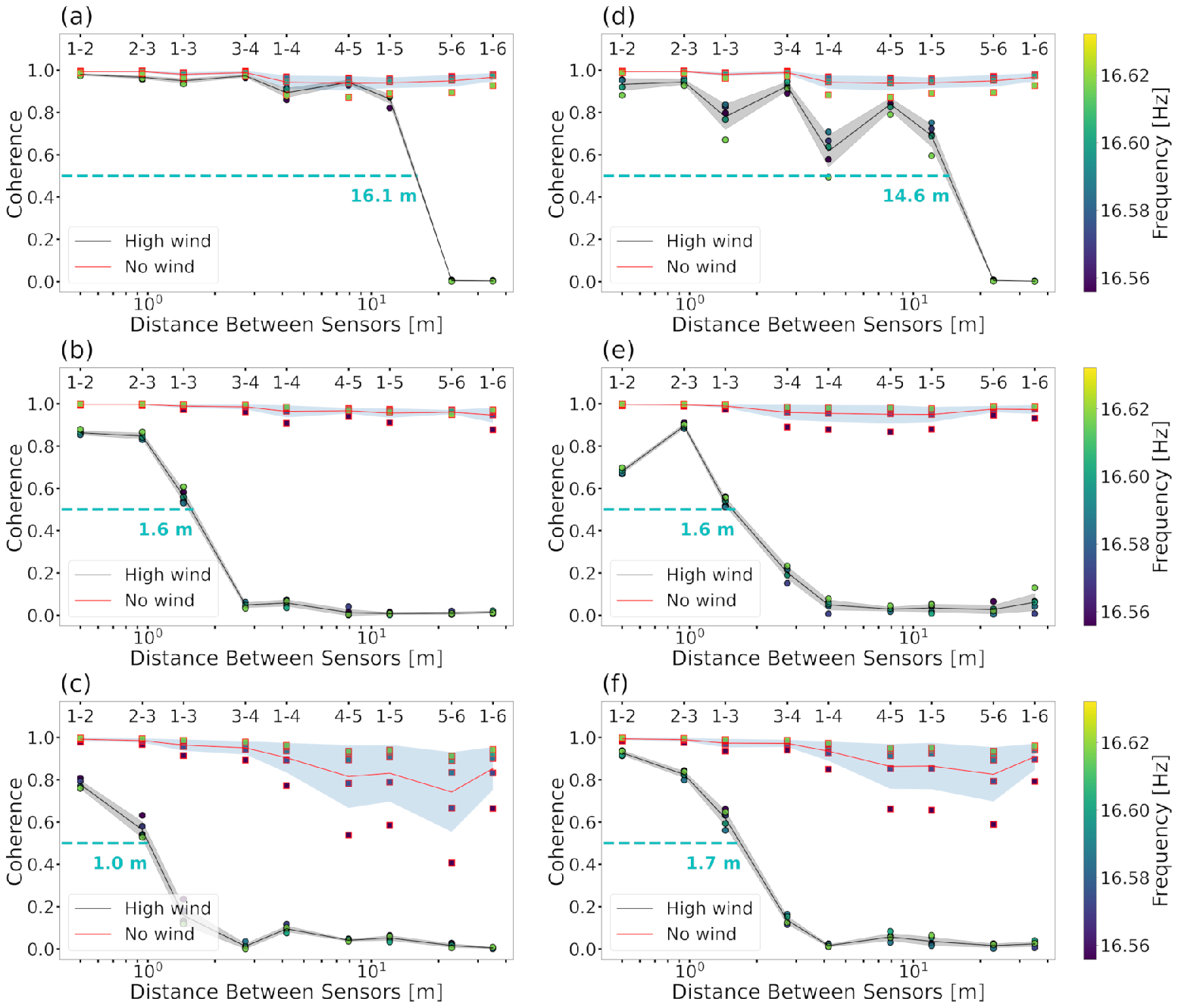}
    \caption{Similar to figures~\ref{Figure_6} and ~\ref{Figure_7} but for 16.55--16.66~Hz frequency band. High wind (black) is compared to non-windy (red) for horizontal components in  S-1 (a) (building), S-2 (b) (shrubs) and S-3 (c) (trees), then similarly for vertical components in (d), (e) and (f) respectively. Solid lines are the averages over the frequency band and shaded areas represent one standard deviation.}
    \label{Figure_8}
\end{figure}

Figure~\ref{Figure_6} shows the coherence lengths plots for the 0.06--0.1~Hz frequency band. This was the lowest frequency and identified as suffering from wind induced noise in the pilot study (figure~\ref{Figure_3}). In the non-windy data the coherence is close to 1 for all seismometer pairs indicating a coherent background seismic noise. In windy conditions the coherence drops rapidly with increasing seismometer separation. This indicates the wind induced seismic noise has a very short coherence length. For the windy data, the shortest lengths with coherence values of 0.5 are highlighted. It can be seen that the coherence length ($\approx 2 m$) is very similar for the horizontal component of the three locations.

Figure~\ref{Figure_7} shows the results for 1.5--2.5~Hz as the second frequency band that  suffers from wind induced seismic noise as identified in the pilot study. In analysis of the dedicated array data, another striking wind driven noise effect was found in a narrow band around 16.6~Hz as shown in figure~\ref{Figure_8}. Our site characterization results show that the coherent signal in this frequency band is generated from a water pump in the study area. Windy coherence lengths for S-2 and S-3 which are in shrubs and amongst tall trees respectively, drop more drastically than those for S-1 which is near the East End-Station building (figure~\ref{Figure_4}) with sparser shrubs and almost no tall trees around its nodes. 

It should be noted that significant low coherence features in the plots are often associated with just one seismometer which suggests wind driven effects can be local. This is thought to be due to different interaction of the different surface objects with wind as observed in similar studies [26]. This effect allows finding the locations with surface objects that vibrate with wind more severely. For example, the pairs that have seismometer 6 of S-1 in common ((a) and (d) in figures~\ref{Figure_6}--\ref{Figure_8}) have significantly lower values. This is a permanent node which has been protected by fences fixed to the ground less than 1~m around the seismometer (G3 in the pilot study). Similarly, figures~\ref{Figure_6}(f), ~\ref{Figure_7}(c) and ~\ref{Figure_7}(f) show that there is local wind driven noise close to seismometer 5 in S-3 based on the lower coherence of the pairs that share it (marked by 4-5, 1-5, 5-6). This seismometer is very close to a tall tree. Excluding seismometers 6 and 5 from S-1 and S-3 respectively (figure~\ref{Figure_6}(d) and~\ref{Figure_6}(f)), leaves similar coherence lengths for the vertical component (figure~\ref{Figure_6}(d)-(f)) for the three sub-arrays in the absence of local effects. This means that a building (S-1), shrubs (S-2) and trees (S-2) are similar in inducing incoherent horizontal and vertical seismic motion due to wind in the primary microseism frequency band. We also note that for the higher frequencies (figures~\ref{Figure_7} and\,~\ref{Figure_8}), shrubs and trees appear to have a larger incoherent wind driven seismic noise than the building. 

\section{Implications for Gravitational Wave Detection}

Feed-forward control is used to suppress ground motion using a seismic array~\cite{Derossa2012}. For these schemes to work, seismic noise must be coherent over the array. The design of such a system generally requires characterization of the site \cite{LIGO_site, VIRGO_site}, design of the array \cite{Coughlin_2016, Badaracco2019} and application of a prediction algorithm. In this paper we consider the Wiener filtering algorithm that has been demonstrated at gravitational wave detector sites \cite{coughlin2014, Coughlin_2019}.

The Wiener filtering approach is used to coherently sum the signals from seismometers in the array, each having the appropriate transfer function applied to maximise the projection of the seismic noise from the seismometers in the array (witness sensors) to the subtraction point (target sensor). Incoherent seismic noise masks the useful coherent signals by adding noise to the coherent sum. The accuracy of a prediction using an array and Wiener filtering can be quantified by the residual between the predicted and the recorded seismic trace at the target location. This metric reduces to:

\begin{equation}
    R(f) = 1 - |\gamma(f)|^2,
    \label{eq2} 
\end{equation}  when we consider a single witness seismometer for simplicity \cite{Coughlin_2019}.

To determine the effect on a gravitational wave detector we imagine various arrays to perform function in various frequency bands of the gravitational wave detector. The primary microseism frequency band is very important for its significance in maintaining stable advanced GW detector operation. Up-converted noise typically contaminates low frequency in the detector sensitivity band which is significant for detection of the inspiral of compact binary coalescenses that allow early warning for multimessenger astronomy \cite{harms, upperlimit, Aasi, helioseis}. Improving vibration isolation in this frequency band with array based feed-forward subtraction will be challenging as the wavelength is of order of a few kilometers, while the coherence length in windy periods is reduced to between a few meters and a few 10s of meters. An array 10s meter in size would have severely reduced angular and wave velocity resolution. An array array km in size would essentially act as a single seismometer with improved signal to noise ratio, while it will have no coherence in windy periods.

For seismic waves between 2 and 2.5Hz the wavelengths are of order of a few hundred meters. In this case seismometer spacing of a few 10s of meters may achieve optimal angular and velocity resolution \cite{Badaracco2019}. If we imagine an array with one witness sensor separated from a target sensor by 35~m, we can refer to figure ~\ref{Figure_7} to see that even in the best case wind reduced the coherence from ~0.9 to ~0.2. This means that with no wind, the array would be able to subtract 81\% of the signal from the target sensor, i.e. reducing effective motion by a factor of 5. This is while in windy periods, the array would only be able to subtract 4\% of the ground motion. This will be a problem for next generation GW detector were NN will still be an issue below 10 Hz \cite{Badaracco2019}.

Above 10~Hz the seismic spectrum at the Gingin site is a complex super-position of wind induced seismic noise and narrow and broad spectral features. This is common at many sites \cite{LIGO_site, VIRGO_site} where machinery like turning motors make narrow spectral features and general movement of people and machinery creates broad spectral features. It is inevitable that there will be spectral features masked by wind noise like the 16.6Hz line shown in figure \ref{Figure_8}. However, our results show that wind induced seismic noise does not pose additional challenges because at this frequency band: 1- the wavelength are short ($\approx 15m$) and the optimum array proposed \cite{Coughlin_2016, Badaracco2019} are generally inside buildings. 2- During high wind at location S-1 close to a building with very sparse shrubs around it, the background seismic noise coming from the electrical pump outside the building is still coherent over lengths longer than the optimum distance proposed for Wiener filtering and the wind driven decrease in the coherence length is the least among the sub-arrays (figure~\ref{Figure_8}(a) and (d)) which means that the building cannot be the source for the incoherent wind induced seismic noise, and 3- the high frequency attenuation effect of the slabs inside the buildings efficiently attenuates the wind induced seismic noise present outside the buildings (figure \ref{Figure_3}(a)).

For the three problematic frequency bands, it may be possible to use more seismometers and have a large array for maximum subtraction in times of coherent noise, while still having a dense small array that will be effective when there is incoherent noise like wind induced noise. However, the extent to which increasing the signal to noise ratio will compensate for the loss of velocity and angular resolution or the changes in the parameters of an optimal array for Wiener filtering must be investigated in more details. Another area of research is to examine the coherency of wind speed and seismic ground motion measured by a dedicated anemometer close to each seismometer and investigate the feasibility of using the wind speed as a proxy to subtract wind induced seismic noise from the seismic records.

\section{Conclusion}

In this paper we quantified the low coherency of wind induced seismic noise for 0.06--20~Hz where advanced GW detector will be most affected. In this band, array based feed-forward and Wiener filtering have been proposed to improve the performance of vibration isolation systems for microseism and Newtonian noise subtraction respectively. The effect of wind induced seismic noise on array performance was quantified by measuring the seismic coherence length in windy and non-windy conditions. We used 3 sub arrays in the vicinity of distinct surface features to identify the effect of such features. The results showed that coherence length decreases sharply with wind for 0.06--0.1~Hz, 1.5--2.5~Hz and 16.55--16.66~Hz frequency bands in the study area. It was demonstrated that for frequencies below 0.1~Hz, wind shortens the coherence length from several hundred meters to 2--$>$35 meters. We find no significant distinct affect in the 3 locations: near a building, in tall tree and in shrubs. At frequencies above 1~Hz the interaction of wind with shrubs and trees shortens the coherence length more than that with a building. Wind induced seismic noise undermines the performance of Wiener Filtering and makes its residual 5 times worse at 2~Hz. This quantification showed that wind induced seismic noise does not pose additional challenge for NN cancellation between 10--20~Hz. This is because optimal arrays to deal with NN at this frequency band have been generally proposed to be deployed inside the  buildings of GW detectors. Such a building not only does not excite challenging low coherent wind driven ground motion at this frequency band, but also its concrete slab attenuates this type of noise  made by shrubs and trees coming form outside the building. The results suggest that wind mechanism that leads to low coherency of seismic noise is a very local effect. Therefore, much more attention should be paid while deploying individual nodes to avoid vibrating surface objects compared to that for conventional applications of array seismology.  Low coherent wind induced seismic noise must be taken into consideration when designing seismic arrays for future gravitational wave detectors because it will affect the number of seismometers required and the ability of the array to resolve seismic wave direction and velocity in windy conditions.

\section{Acknowledgement}

This project was supported by ARC Center of Excellence for Gravitational Wave Discovery (OZGRAV), LIEF grant and CSIRO. The authors would like to also thank Vladimir Bossilkov, John More and Steve Key for their assistance in the field deployment of the seismometers.

\textbf{References}


\providecommand{\newblock}{}

\end{document}